\documentclass[%
 reprint,
 amsmath,amssymb,
 aps,
showpacs,
]{revtex4-1}

\usepackage{graphicx}
\usepackage{dcolumn}
\usepackage{bm}

\begin{document}

\preprint{APS/123-QED}

\title{Constraints on Lorentz Invariance Violation using INTEGRAL/IBIS observations of GRB041219A}

\author{P. Laurent}
\affiliation{APC Laboratory, CEA/IRFU, Universit\'e Paris Diderot, CNRS/IN2P3, Observatoire de Paris, 10, rue Alice Domon et L\'eonie Duquet,
75205 Paris Cedex 13, France}
 \email{plaurent@cea.fr}
\author{D. G\"otz}
\affiliation{AIM (UMR 7158 CEA/DSM-CNRS-Universit\'e Paris Diderot) Irfu/Service d’Astrophysique, Saclay, F-91191 Gif-sur-Yvette Cedex, France}

\author{P. Bin\'etruy}
\affiliation{APC Laboratory, Universit\'e Paris-Diderot, CNRS/IN2P3, CEA/IRFU, Observatoire de Paris, 10, rue Alice Domon et L\'eonie Duquet,
75205 Paris Cedex 13, France}

\author{S. Covino}
\affiliation{ INAF – Osservatorio Astronomico di Brera, Via E. Bianchi 46, 23807 Merate (LC), Italy}

\author{A. Fernandez-Soto}
\affiliation{Instituto de Fisica de Cantabria, CSIC-Universidad Cantabria, Avenida de los Castros s/n, 39005 Santander, Spain}

\date{\today}

\begin{abstract}
One of the experimental tests of Lorentz invariance violation
is to measure the helicity dependence of the propagation
velocity of photons originating in distant cosmological obejcts.
Using a recent determination of the distance of the Gamma-Ray
Burst GRB 041219A, for which a high degree of polarization is observed in the
prompt emission, we are able to improve by 4 orders of magnitude the existing constraint on 
Lorentz invariance violation, arising from the phenomenon of vacuum birefringence.
\end{abstract}


\maketitle

\section{\label{sec:level1}Introduction}

On general grounds one expects that the two fundamental theories of 
contemporary physics, the theory of General Relativity and the quantum theory 
in the form of the Standard Model of particle physics, can be unified at the 
Planck energy scale.  This unification requires to quantize gravity, which 
leads
to very fundamental difficulties. One is related with the energy of the 
fundamental vacumm state. Another one is the status of Lorentz invariance: the fuzzy 
nature of space time in quantum gravity may lead to violations of this fundamental symmetry.   
For the last two decades theoretical studies and experimental searches of 
Lorentz Invariance Violation (LIV) have received a lot of attention 
\cite[see e.g. the reviews by][]{mattingly05,jacobson06,liberati09}. Possible 
consequences of LIV are energy and helicity dependent photon 
propagation velocities. The energy dependence can be constrained by recording 
the arrival times of photons of different energies emitted by distant objects 
at approximately the same time \cite{amelino98}, e.g. during a Gamma-Ray Burst 
\cite{abdo09} or a flare of an Active Galactic Nucleus \cite{aharonian}. On 
the other hand, the helicity dependence can be constrained by measuring how 
the polarization direction of an X-ray beam of cosmological origin changes as 
function of energy \cite{gambini}. 

Recently, an upper limit on the helicity dependence of photon propagation has been set using INTEGRAL/SPI observations of the polarization of the Crab nebulae \citep{maccione08}. In this paper, we derive much stronger constraints from a polarization measurement of a Gamma-Ray Bursts (GRB), a source at cosmological distance.

To date only a few polarization measurements are available for GRBs. 
\citet{coburn03} 
reported a high degree of polarization, $\Pi = 80\% \pm 20\%$,
for GRB 021206.
However, successive reanalysis of the same data set could not confirm this claim, reporting
a degree of polarization compatible with zero \citep{rutledge04,wigger04}.
\citet{willis05} 
reported a strong polarization
signal in GRB 930131 ($\Pi > 35\%$) and GRB 960924 ($\Pi > 50\%$), but
this result could not be statistically constrained.
GRB 041219A was detected by the INTEGRAL Burst Alert System (IBAS; \citealt{ibas}), and 
is the longest and brightest GRB localized by INTEGRAL \citep{integral} to date \cite{vianello08}.
\citet{mcglynn07}, using the data of the INTEGRAL spectrometer \citep[SPI;][]{spi},  reported a high degree
of polarization of the prompt emission 
($\Pi=68\pm29\%$) for the brightest part of this GRB. 
Also,  \citet{gotz09}  have performed a similar measurement, using this time the INTEGRAL/IBIS telescope, reporting variable polarization  properties of the burst all along its duration. In this paper, we reuse these data in order to measure the polarization in two energy bands and check for a shift in the
polarization angle as a possible effect of Lorentz Invariance Violation. 

In section \ref{sec:level1}, we will present the INTEGRAL/IBIS observations of polarization during the burst, and the recent measure of the burst distance we obtained thanks to far infrared observations of its host galaxy. In section \ref{sec:level2}, we will derive constraints on LIV from these observations and we will conclude in section \ref{sec:level3}.

\section{\label{sec:level1}Observations of GRB041219A}

\subsection{INTEGRAL/IBIS observations}

Thanks to its two position sensitive detector layers ISGRI \citep{isgri} (made of CdTe crystals and sensitive in 
the 15--1000 keV energy band), and PICsIT \citep{picsit} (made of CsI bars and sensitive in the 200 keV--10 MeV energy band), IBIS can be used as a Compton polarimeter \citep{lei97}. The concept behind a Compton polarimeter is
the polarization dependency of the differential cross section for Compton scattering 

\begin{eqnarray}
\frac{d\sigma}{d\Omega} = \frac{r_{0}^{2}}{2}\left(\frac{E^{\prime}}{E_{0}}\right)^{2}\left(\frac{E^{\prime}}{E_{0}}+\frac{E_{0}}{E^{\prime}}-2 \sin^{2}\theta \cos^{2}\phi \right)
\end{eqnarray}
where $r_{0}^{2}$ is the classical electron radius, $E_{0}$ the energy of the incident photon, $E^{\prime}$
the energy of the scattered photon, $\theta$ the scattering angle, and $\phi$ the azimuthal angle relative
to the polarization direction. Linearly polarized photons scatter preferentially perpendicularly to the incident
polarization vector. Hence by examining the scatter angle distribution of the detected photons

\begin{eqnarray}
N(\phi)=S[1+a_{0}\cos(2(\phi-\phi_{0}))],
\label{eq:azimuth}
\end{eqnarray}

one can derive the polarization angle $PA =  \phi_{0} - \pi /2 + n \pi$ and the polarization fraction 
$\Pi= a_{0}/a_{100}$, where $a_{100}$ is the amplitude expected for a 100\% polarized source, derived
by Montecarlo simulations  of the instrument.

To measure the polarization, we followed the same procedure described in \citet{forot08} that allowed to successfully detect a polarized signal from the Crab nebula. One important difference, anyway, is that the Crab measurement was integrated over several observation periods and a long
time ($>1$Ms), while this measurement here integrates over only a few seconds. So we have checked if  instrumental azimuthal variations, potentially washed out by the long Crab observation, could be important in the GRB case. To do so, we have simulated ten GRB-like observations using data from the INTEGRAL Payload Ground Calibrations campaign. This campaign, made with the full flight model of the telescope, was done with standard radioactive non polarized sources. We use in this work the calibrations made with a $^{113}$Sn source with a peak energy of 392 keV, the closest to the energy bands  we are considering here. To simulate GRBs, we divide the data set in ten subsets containing each a similar number of Compton events as in the observations we show in figure  \ref{fig:evol}. In all these subsets, we found systematics at a level up to $15 \%$ maximum, with a polarization angle around 40$^{\circ}$.
We made another tests considering spurious events, as described in  \citet{forot08}, and again we found a systematic azimuthal variation of less than $9 \%$, with a polarization angle of $180^\circ$. Both value are far from the ones we derived below for the source and give confidence that our results are not due to instrumental azimuthal variations, even if such variations effectively exist.

To derive the Gamma-Ray Burst flux as a function of $\phi$, the Compton photons were divided in 6 bins of 30$^{\circ}$ as a function of the azimuthal scattering angle. To improve the signal-to-noise ratio in each bin, we took advantage of the $\pi$-symmetry of the differential cross section (see Eq. \ref{eq:azimuth}), i.e. the first bin contains the photons with $0^{\circ}<\phi<30^{\circ}$ and $180^{\circ}<\phi<210^{\circ}$, etc. The derived detector images were then deconvolved to obtain sky images, where the flux of the source in each bin is measured by fitting the instrumental  PSF to the source peak. We finally fitted the polarigrams to Eq. \ref{eq:azimuth} using a least squares technique (see Fig. \ref{fig:pola}) in order to derive $a_{0}$ and $\phi_{0}$, and the errors on the parameters are dominated 
by the statistics of the data points. 

We computed the scatter angle distribution into two energy band in order to detect a possible polarimetric angle shift with energy, reminiscent of a possible LIV effect. The two bands were chosen to be [200-250 keV] and [250-325 keV] where the source has merely the same signal to noise ratio. In figure  \ref{fig:evol}, we show the measured evolution of the polarimetric angle shift between these two energy bands, along the burst duration. These shifts are all consistent with zero with a mean value of $21^\circ \pm 47^\circ$.

Also, as an exemple, we show in figure  \ref{fig:pola} and figure \ref{fig:contour1}, the portion of the GRB light curve where the polarimetric signal was strong, that is starting at 01:47:02 U.T. until 01:47:12 U.T. (P9 interval in \citet{gotz09}).  A modulated signal is seen in the two energy bands, corresponding to $\Pi = 55 \%$ for the first band, and  $\Pi = 82 \%$  for the second ones (see fig.\ref{fig:contour1} for error contours). To evaluate the goodness of our fits, we computed the chance probability (see Eq. 2 in \citealt{forot08}) that our polarigrams are due to an unpolarized
signal, and reported these values in Fig. \ref{fig:pola}. The corresponding polarimetric angles were  $PA = 80^{+26}_{-28}$ deg. and  $PA = 45^{+38}_{-40}$ deg., that is consistent at the $2 \sigma$ level (see  figure \ref{fig:contour1}). Propagation of errors  gives an upper limit of $68^\circ$ at the $90 \%$ level, for a possible phase shift.

\begin{figure}[ht]
\includegraphics[width=9cm,angle=0]{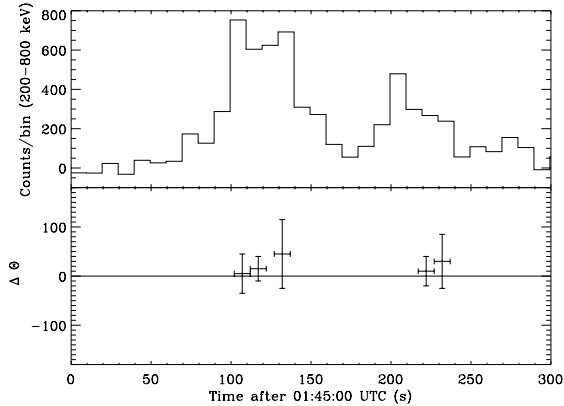}
\caption{\label{fig:evol} Evolution during the burst duration of the polarimetric angle shift measured between the [200-250 keV] and in the [250-325 keV] energy range. The mean value, $21^\circ \pm 47^\circ$, is consistent with zero.}
\label{fig:evol}
\end{figure}

\begin{figure}[ht]
\includegraphics[width=9cm,angle=0]{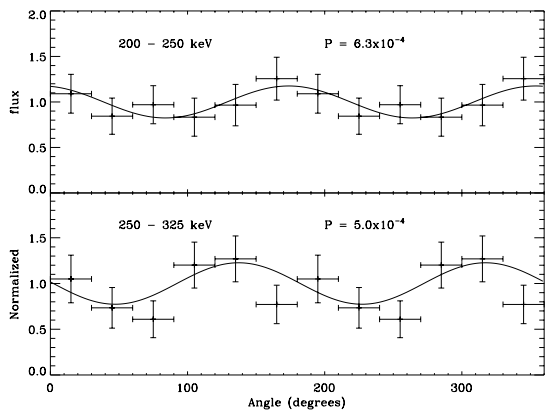}
\caption{\label{fig:pola} Scatter angle distribution of GRB041219A in the 200-250 keV and in the 250-325 keV energy band during the P9 time interval  (see text). These distributions give the source count rate by azimuthal angle of the Compton scattering, and are consistent with a highly polarized signal. The chance probability of a non-polarized signal is reported in each panel. The polarization angles derived from these distributions are consistent within 68$^\circ$ (see text).}
\label{fig:pola}
\end{figure}

\begin{figure}[ht]
\includegraphics[trim = 2.5cm 13cm 1cm 5cm, clip,width=9cm,angle=0]{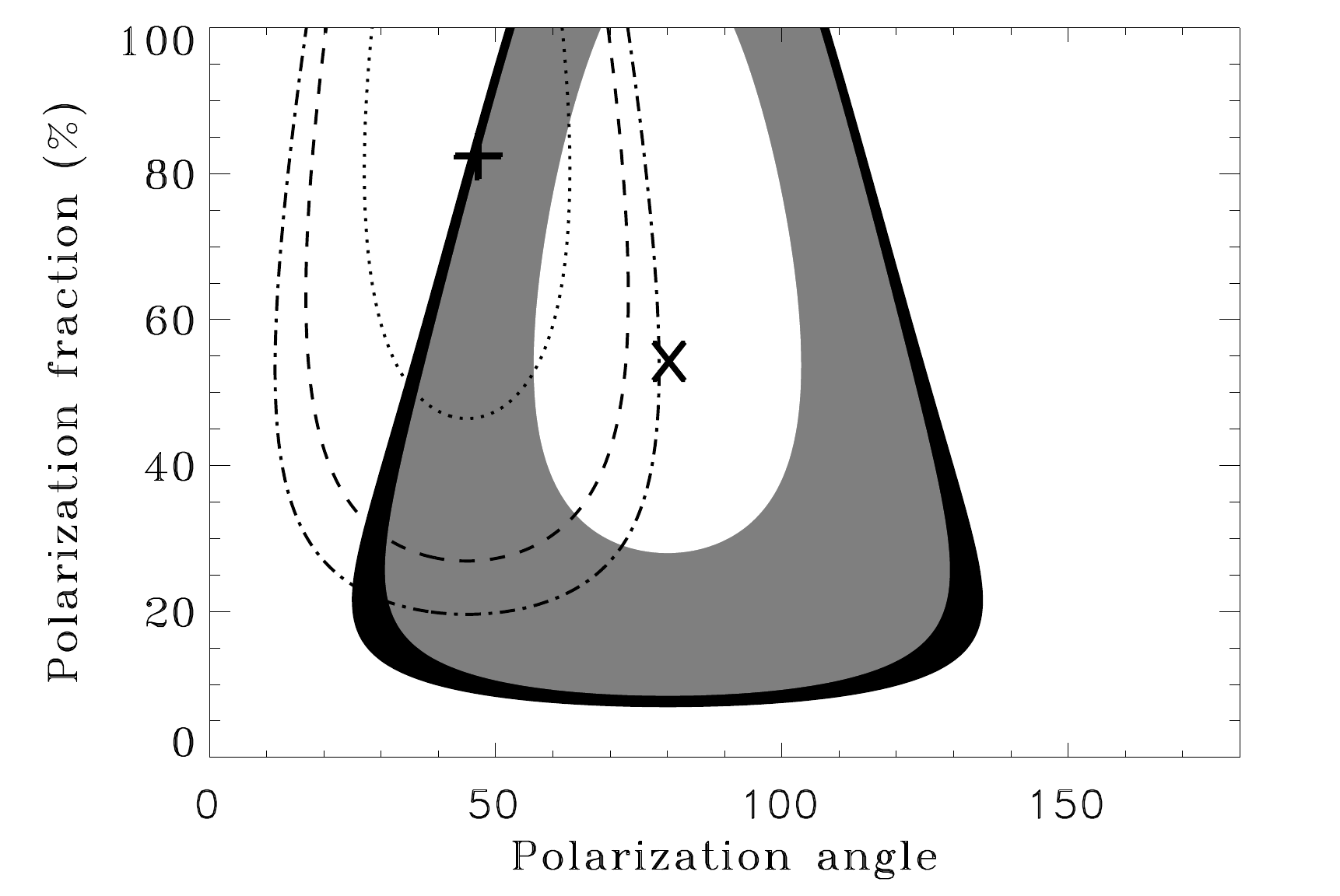}
\caption{\label{fig:contour1} Contour plot of the composed error on polarisation angle and polarisation fraction for the two energy bands, during the P9 time interval. The X cross shows the best fit position for the first energy band. Contour levels are at the 67, 90, and 95 \% level (from white toward black fill). The + cross and dotted, dashed, and dash-dotted lines show the best fit parameters obtained for the second energy band, consistent at $95 \%$ with the first energy band best fit.}
\label{fig:contour1}
\end{figure}

\subsection{Distance determination}

\citet{gotz11} performed deep infra-red imaging of the GRB region using the WIRCam instrument at the 3.6 m Canadian French Hawaiian Telescope (CFHT) at Mauna Kea. Thanks to multi-band ($YJHK_{s}$) imaging they were able to identify the host galaxy of GRB 041219A, and to give an lower limit to its
photomeric redshift of $z=0.02$, at the $90 \%$ confidence level. This implies a luminosity distance of 85 Mpc, assuming standard cosmological parameters ($\Omega_{m} = 0.3$, $\Omega_{\lambda} = 0.7$, and $H_{0} = 70$ km/s/Mpc).  

\section{\label{sec:level2}Constraints on Lorentz Invariance Violation}

On general grounds, Lorentz violating operators of dimension $N=n+2$  
modify the standard dispersion relations $E^2 = p^2 + m^2$ by terms of the 
order of $f_n p^n/M_{Pl}^{n-2}$ where $M_{Pl}$ is the reduced Planck scale
($\approx 2.4~ 10^{18}~$ GeV), used as a 
reference scale since LIV is expected to arise in the quantum regime of gravity.
In order to account for the severe limits on LIV, one therefore usually only 
considers operators of dimension greater or equal to 5 which provide 
corrections which are tamed by at least one inverse Plank scale.  
 
\subsection{dimension 5 operators}

If we restrict our attention to pure electrodynamics, there is a single term of
dimension 5 which gives corrections of order $p^3/M_{Pl}$ and is compatible 
with gauge invariance and rotational symmetry \cite{myerspospelov}:
\begin{equation}
{\cal L} = {\xi \over M_{Pl}} n^\mu F_{\mu\nu} n^\rho \partial_\rho 
\left(n_\sigma\tilde F^{\sigma\nu} \right)\ ,
\label{dim5}
\end{equation}
where $n^\mu$ is a 4-vector that characterizes the preferred frame and 
$\tilde F^{\mu\nu}\equiv \frac{1}{2} \epsilon^{\mu\nu\rho\sigma}
F_{\rho\sigma}$.
The uniqueness of this term makes the analysis somewhat model-independent 
(see however below).

The light dispersion relation is given by ($E=\hbar \omega$ and $p= \hbar k$):
\begin{eqnarray}
\omega^2 = k^2 \pm \frac{2\xi  k^3}{M_{Pl}}\equiv \omega^2_\pm \ .
\label{eq:one}
\end{eqnarray}
where the sign of the cubic term is determined by the chirality (or circular 
polarization) of the photons, which leads to a rotation of the polarization  
during the propagation of linearly polarized photons. This effect is known as 
vacuum birefringence.

Since we have the approximative relation:
\begin{eqnarray}
\omega_\pm = | p| \sqrt{1 \pm \frac{2\xi  k}{M_{Pl}}} \approx |k|  (1 \pm 
\frac{\xi  k}{M_{Pl}})\ ,
\label{eq:two}
\end{eqnarray}
the direction of polarization rotates during propagation along a distance $d$
by an angle:
\begin{eqnarray}
\Delta\theta(p) = \frac{\omega_+(k) - \omega_-(k)}{2} d  \approx \xi 
\frac{ k^2 d}{2M_{Pl}} \ .
\label{eq:three}
\end{eqnarray}
For GRB041219A, if we set $\Delta\theta(k) = 47^\circ$, derived from the 
measures we made along the burst duration, and the lower limit luminosity 
distance reported above of $d = 85$ Mpc = $2.6 ~10^{26}$ cm, corresponding to 
z=0.02, we get an upper limit on the vacuum birefringence effect:

\begin{eqnarray}
\xi < \frac{2M_{Pl} \Delta\theta(k)}{(k_2^2-k_1^2) d} \approx 1.1\ 10^{-14}
\label{eq:four}
\end{eqnarray}

\subsection{dimension 6 operators}

Although the dimension 5 operator (\ref{dim5}) is unique, it is physicaly 
relevant only if 
there does not appear operators of lower dimensions. In the general case
however, radiative corrections induced by dimension 5 operators lead to
dimension 3 operators through quadratic divergences which contribute an extra 
factor $\Lambda^2$, where $\Lambda$ is an ultraviolet cut-off of order 
$M_{Pl}$.
Barring extreme fine tuning, one needs a symmetry argument to cancel such 
terms. 
Supersymmetry appears to be the only symmetry which cancels such 
contributions \cite{NibbelinkPospelov}. 
It is true that supersymmetry is broken in nature but, in the case where it is 
softly broken, quadratic divergences contribute a factor $M_S^2$, where $M_S$,
the scale of supersymmetry breaking, is of the 
order of a TeV$^2$, thus providing an extra factor $(M_S/M_{Pl})^2 \sim
10^{-30}$ compared to untamed dimension 3 operators. 

Unfortunately, the dimension 5 operator of (\ref{dim5}) is not compatible with
supersymmetry \cite{NibbelinkPospelov}. We therefore have to resort in this 
case to dimension 6 operators.  
We thus assume that the light dispersion relation is given by:
\begin{eqnarray}
\omega_\pm^2 = k^2 \pm \frac{\xi  k^4}{M_{Pl}^2} \ .
\label{eq:one}
\end{eqnarray}
We can derive then the approximate relation:
\begin{eqnarray}
\omega_\pm = | k| \sqrt{1 \pm \frac{\xi  k^2}{M_{Pl}^2}} \approx |k|  (1 \pm \frac{\xi  k^2}{2 M_{Pl}^2})
\label{eq:two}
\end{eqnarray}

which implies that the direction of polarization rotates during propagation of:

\begin{eqnarray}
\Delta\theta(k) = \frac{\omega_+(k) - \omega_-(k)}{2} d  \approx \xi \frac{ k^3 d}{M_{Pl}^2}
\label{eq:three}
\end{eqnarray}

 Again, for GRB041219A, if we set $\Delta\theta(k) = 47^\circ$ and the luminosity distance $d = 85$ Mpc = $2.6 ~10^{26}$ cm, corresponding to z=0.02, we get an upper limit on the vacuum birefringence effect:

\begin{eqnarray}
\xi < \frac{2M_{Pl}^2 \Delta\theta(k)}{(k_2^3-k_1^3) d} \approx 2.6~ 10^{8} 
\label{eq:four}
\end{eqnarray}
This is still too large an upper bound to be really constraining since one 
expects couplings at most of order one.

\section{\label{sec:level3}Conclusions}
Using a recent determination of the distance of GRB041219A \cite{gotz11} for 
which a high degree of polarization is observed \cite{mcglynn07,gotz09}, we 
were able to increase by 4 orders of magnitude the existing constraint on 
Lorentz invariance violations, arising from birefringence. Turned into a
constraint on the coupling $\xi$ of dimension 5 Lorentz violating interactions,
that is of corrections of order $k/M_{Pl}$ to the dispersion relations, this 
gives the very stringent constraint $\xi < 10^{-14}$. Most presumably, this 
means that such operators are vanishing, which might point towards a symmetry 
such as supersymmetry in action. In that case, the pressure is on the next 
corrections of order $(k/M_{Pl})^2$ corresponding to operators of dimension 6. 
We showed that, although astrophysical constraints are not yet really 
constraining, they are getting closer to the relevant regime ($\xi$ of order 
$1$ or smaller).
Our result can be compared to the limits derived using the possible energy dependence of the group velocity of photons in distant GRBs derived with Fermi \cite{abdo09}. However, written in the same way as we did in this work, this limit was $\xi<0.8$ that is much less constraining.

\nocite{*}

\acknowledgements{ISGRI has been realized and maintained in flight by CEA-Saclay/Irfu with
the support of CNES. Based on observations with INTEGRAL, an ESA project with instruments and science data centre funded by ESA member states (especially the PI countries: Denmark, France, Germany, Italy, Switzerland, Spain), Czech Republic and Poland, and with the participation of Russia and the USA. }

\end{document}